# PROGRESS ON THE 140 KV, 10 MEGAWATT PEAK, 1 MEGAWATT AVERAGE POLYPHASE QUASI-RESONANT BRIDGE, BOOST CONVERTER/MODULATOR FOR THE SPALLATION NEUTRON SOURCE (SNS) KLYSTRON POWER SYSTEM [1]

William A. Reass, James D. Doss, Robert F. Gribble, Michael T. Lynch,
and Paul J. Tallerico
Los Alamos National Laboratory, P.O. Box 1663, Los Alamos, NM 87544

*Abstract*

This paper describes electrical design and operational characteristics of a zero-voltage-switching 20 kHz polyphase bridge, boost converter/modulator for klystron pulse application. The DC-DC converter derives the buss voltages from a standard 13.8 kV to 2300 Y substation cast-core transformer. Energy storage and filtering is provided by self-clearing metallized hazy polypropylene traction capacitors. Three "H-Bridge" IGBT switching networks are used to generate the polyphase 20 kHz transformers primary drive waveforms. The 20 kHz drive waveforms are chirped the appropriate duration to generate the desired klystron pulse width. PWM (pulse width modulation) of the individual 20 kHz pulses is utilized to provide regulated output waveforms with adaptive feedforward and feedback techniques. The boost transformer design utilizes amorphous nanocrystalline material that provides the required low core loss at design flux levels and switching frequencies. Resonant shunt peaking is used on the transformer secondary to boost output voltage and resonate transformer leakage inductance. With the appropriate transformer leakage inductance and peaking capacitance, zero-voltage-switching of the IGBT's is attained, minimizing switching losses. A review of these design parameters and a comparison of computer calculations, scale model, and first article results will be performed.

## 1 MAJOR COMPONENTS

The simplified bock diagram of the converter/modulator system is given in Figure 1. This system economizes costs by the utilization of many standard industrial and utility components. The substation is a standard 13.8 kV to 2300 Y cast-core transformer with passive harmonic traps and input harmonic chokes. These components are located in an outdoor rated NEMA 3R enclosure that does not require secondary oil containment or related fire suppression equipment. An SCR regulator is helpful to optimize the IGBT switching parameters. The SCR regulator also accommodates incoming line voltage variations and other voltage changes resulting from transformer and trap impedances, from no-load to full-load. The SCR regulator also provides the soft-start function. The SCR regulator provides a nominal +/- 1500 Volt output to the energy storage capacitors. The energy storage capacitors are self-clearing metallized hazy polypropylene traction motor capacitors. As in traction application, these capacitors are hard bussed parallel. These capacitors do not fail short, but fuse or "clear" any internal anomaly. At our capacitor voltage rating (2 kV) there has not been a recorded internal capacitor buss failure. In this application, as in traction motor application, the capacitor lifetime is calculated to be 1e9 hours. Insulated Gate Bipolar Transistors (IGBT's) are configured into three "H" bridge circuits to generate a three phase 20 kHz square wave voltage drive waveform. The IGBT's are "chirped" the appropriate duration to generate the high voltage klystron pulse, typically 1.3 mS. Due to the IGBT switching power levels, currents, and frequencies involved, low inductance busswork and bypassing is of paramount importance. The boost transformers utilize amorphous nanocrystalline material that has very low loss at the applied frequency and flux swing. Operating at 20 kHz and about 1.6 Tesla bi-directional, the core loss is about 1.2 watts per pound in our application, or 320 W per core. Each of the "C" cores (one for each phase) weigh 260 lbs. and has a 3.5" by 5" post. By appropriately spacing the secondary from the primary, the transformer leakage inductance can be resonated with secondary shunt peaking capacitors to maximize voltage output and tune the IGBT switch current to provide "zero-voltage-switching". The zero-voltage-switching is provided when the IGBT gate drive is positive, but reverse transformer primary circulating current is being carried by the IGBT internal freewheel diode. We have tuned for about 4 uS of freewheel current before the IGBT conducts in the normal quadrant. This tuning provides for about 15% control range for IGBT pulse width modulation (PWM). The ability of IGBT PWM of the active klystron voltage pulse permits adaptive feedback and feedforward techniques with digital signal processors (DSP's) to regulate and provide "flat" klystron voltage pulses, irrespective of capacitor bank "start" voltage and related droop. Line synchronization is not absolutely required as the adaptive DSP can read bank voltage parameters at the start of each pulse and calculate expected droop. A standard six-pulse rectification circuit is used with a "pi-R" type filter network. The diodes are fast recovery ion-implanted types that are series connected with the appropriate compensation networks. The diodes have the second highest power loss (after the IGBT's) and are forced oil cooled. The filter network must attenuate the 120 kHz switching ripple and have a minimal stored energy. The stored energy is wasted energy that must be dissipated by the klystron at the end of each pulse. With the parameters we have chosen, the ripple is very low (~300 volts) and the klystron fault energy (in an arc-down) is about 10 joules. Even if the IGBT's are not turned off, the transformer resonance is out of tune because of the fault condition, and little difference in klystron fault energy is realized. If the IGBT's fail short, through the transformer primary winding, the boost transformer will saturate in about 30 uS, also limiting any destructive faults to the klystron. In a faulted condition, the klystron peak fault current is about twice nominal, with low dI/dT's.

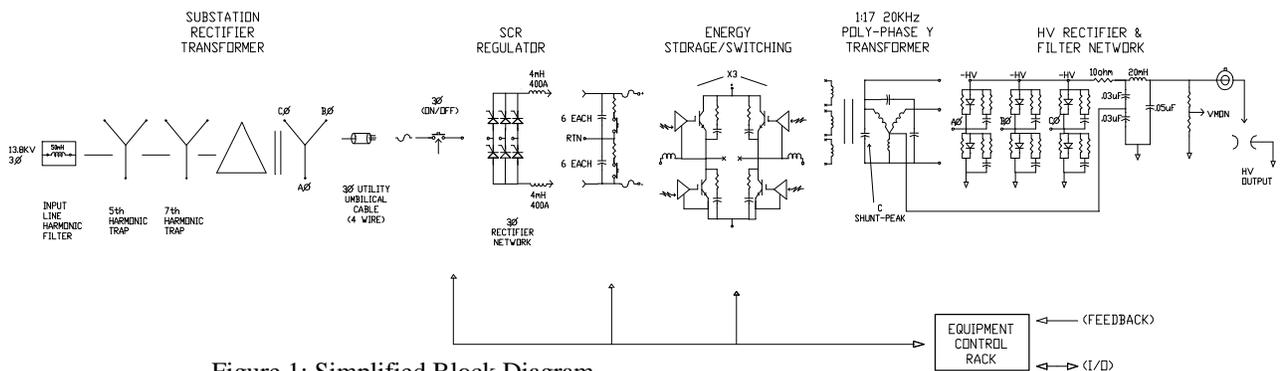

Figure 1: Simplified Block Diagram.

___

[1] Work supported by The U.S. Department of Energy

## 2 MODELING

The complete electrical system of the converter/modulator system has been modeled in extreme detail. This includes design studies of the utility characteristics, transformer and rectification methodology, IGBT switching losses, boost transformer parameters, failure modes, fault analysis, and system efficiencies. Various codes such as SCEPTRE, MicrocapIV, Flux2D, and Flux3D have been used to complete these tasks. SCEPTRE has been primarily used to examine IGBT and boost transformer performance in great detail to understand design parameters such as switching losses, IGBT commutation dI/dT, buss inductance, buss transients, core flux, core flux offset, and transformer Eigen frequencies. Flux2D and Flux3D have been used to examine transformer coupling coefficients, leakage inductance, core internal and external flux lines, winding electric field potentials, and winding field stresses. The Flux2D and Flux3D were particularly useful to examine transformer secondary winding profiles that gave the desired coupling coefficients with minimized electrical field stresses. MicrocapIV has been used to examine overall design performance of the system. This includes the utility grid parameters such as power factor, line harmonics, and flicker. We have optimized the design to accommodate the IEEE-519 and IEEE-141 standards. The MicrocapIV uses simplified switch models for the IGBT's, which does not accurately predict their losses. However, the code has been very useful to examine tradeoffs of circuit performance with the lumped elements such as the boost transformer, shunt peaking capacitance, the filter networks, and the input energy stores. MicrocapIV is also adept at making parametric scans to determine component sensitivities and tolerances. Comparisons between the SCEPTRE and MicrocapIV codes show no significant differences in the system operational performance such as switching currents, switching voltages, and output voltage parameters. A typical output of MicrocapIV, Figure2, shows the zero-voltage-switching of a pair of IGBT's in the top waveform as compared to the transformer primary winding current in the middle waveform.

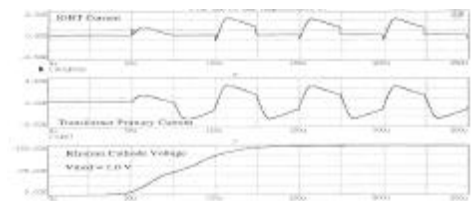

Figure 2: Details of Waveforms.

The negative IGBT current before turn-on is the freewheel diode current. This negative current is of paramount importance to provide a loss-less turn-on of the IGBT, excessive PWM can cause the IGBT to fall out of synchronization with the winding current. Limits of PWM are necessary to ensure this condition does not happen, causing over-heating of the IGBT's. A detail of the klystron voltage rise time, about 75 uS, is shown at the bottom. With the appropriate optimization, good klystron pulse fidelity can be obtained as shown in Figure 3. The model shows the resulting droop from the capacitor energy stores.

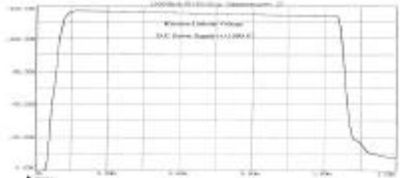

Figure 3   Klystron Cathode Voltage.

## 3   SCALE MODEL RESULTS

A scale model has been built and has been used extensively to prove various design concepts. As electronic control subsystems become available, they can be tested and debugged with the scale model. Figure 4 shows the first results we obtained with our adaptive power supply feedback and feedforward system. The lower most trace shows the capacitor bank voltage droop of 20%, five times worse than the SNS bank design (~4% droop). The scale model shows a droop that follows the capacitor voltage (as can be expected) without any adaptive controls. With adaptive control, the flattop performance is considerably improved as well as having a better rise time. Further efforts to optimize the loop parameters should be able to provide a flat output voltage pulse.

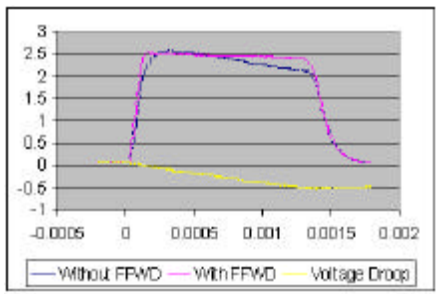

Figure 4:  Adaptive Feedfoward/Feedback

## 4 PROJECT STATUS

The status of the first article is moving along well. The construction efforts are nearing completion with only a few subassemblies needed before final integration. The oil tank basket is shown in Figure 5. The high-voltage rectifier assemblies are in the lower portion of the figure. The IGBT switch boards can be seen on top of the converter/modulator in their "edge" connectors. The oil tank and safety enclosure is depicted in Figure 6. Most prominent is the cooling panel for the IGBT's and oil/water heat exchanger. The enclosure is an integral part of a ground plane (also noted in the picture) to minimize any radiated noise. The converter/modulator is completely (electrically) sealed at all points. Unfortunately, vendor delays of the SCR phase control rack will not permit full average power operation until December 00 or January 01. We will therefore use a pair of 10 kW power supplies for initial debug operations and testing to full output voltage, current, and pulse length, but at a reduced pulse rate.

Figure 5:  Oil Tank Basket Assembly

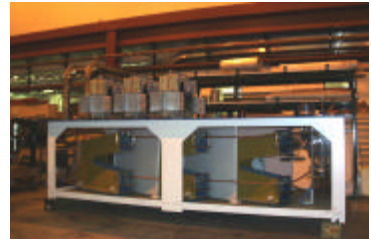

Figure 6:  Oil Tank & Safety Enclosure

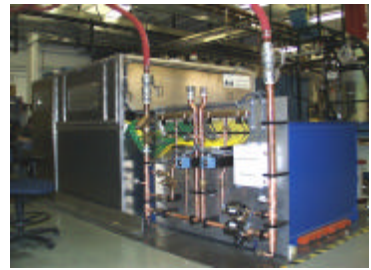

## 5   CONCLUSION

The converter/modulator will be demonstrating many new technologies that will hopefully revolutionize long-pulse klystron modulator design. These items include self-clearing capacitors in a modulator application, large amorphous nanocrystalline cores, high-voltage and high-power polyphase quasi-resonant DC-DC conversion, and adaptive power supply control techniques. These design and construction efforts will hopefully come to closure in less than a year after conception. Design economies should be realized by the use of industrial traction motor components (IGBT's and self-clearing capacitors) and standard utility cast-core power transformers. The modular design minimizes on-site construction and a simplified utility interconnection scheme further reduces installation costs. The design does not require capacitor rooms and related crowbars. Design flexibility is available to operate klystrons of different voltages by primarily changing the boost transformer turns ratio. All testing of individual full-scale components and scale model results all agree to date with modeling efforts that indicate the design technology will be imminently successful.